# Investigation of glow discharge plasma energy distribution using a gridded energy analyzer considering plasma-facing materials related processes

Ali Masoudi, Davoud Iraji[*]

*Department of Physics and Energy Engineering, Amirkabir University of Technology, Tehran, Iran*

[*]E-mail: Iraji@aut.ac.ir

**Abstract**

Considering the effects of glow discharge plasmas on plasma-facing materials and the applications such as coating, cleaning and surface treating, this work has been done to investigate the energy of ions of dc glow discharge plasmas. On the way towards this goal, a plasma chamber has been simulated via COMSOL Multiphysics software. Then, a gridded energy analyzer has been simulated and designed. In the next step, the analyzer has been constructed and tested to measure the energy of plasma ions. The devise contains a grid which is negatively biased to the same potential as the glow discharge cathode electrode. It discriminates plasma ions based on their energies, which are accelerated due to sheath potential drop before colliding with the cathode. The obtained energy distribution function from experiments has been compared to that of simulated plasma. The experimental results show that there are different groups of ions each in local thermal equilibrium in dc glow discharge plasmas.

**Keywords:** Plasma-facing materials, Glow discharge plasma, Energy distribution function, Gridded energy analyzer

## I. Introduction

Nowadays, GDP[1] is a versatile method for surface cleaning and treatments. Some materials and alloys need to be pre-treated to become strong in adhesion for industrial and biological applications such as orthopedic implantation [1], [2], [3]. Improvement of surface properties of semiconductors and metals and ion implantation are other applications of GDP [4]. Stainless steels can be nitrided by GDP in order to improve their wear, mechanical and corrosion resistance [5]. GDP is a common way to remove dusts and impurities from PFMs[2] inside Tokamak vessels [6], [7]. Argon GDC[3] is utilized for hydrogen isotope removal from PFMs in fusion devices [8]. GDC with a combination of argon gas and $H_2$ has recently been used for impurity removal from PFMs in ADITYA-U Tokamak [9]. Xi Yi et al., compared the efficiency of argon plasma with nitrogen plasma in GDC and stated that it is essential to consider the cleaning damages in case of high energy cleanings [10].

---

[1] Glow discharge plasma
[2] Plasma-Facing Materials
[3] Glow discharge cleaning

In GDC process, one or more anodes are placed in the plasma chamber and the chamber wall acts as cathode and is bombarded by ions [11], [12]. The energy and density of plasma ions impinging the surface are of significant importance and a proper flux is required to well treat the PFMs surface while avoiding any considerable damages to them [13]. Therefore, it is necessary to know EDF[4] of ions and to find out if the experimental energy distribution of ions be desire for GDC [14]. Several methods have been developed on the way of the determination of the EDF of plasma ions and electrons and instruments such as Longmuir Probes, Mach probes and Faraday cups provide useful information about the flux and the energy of the ions [15], [16], [17], [18], [19], [20]. The EDF of ions and electrons of the bulk plasma is important but in cases that the treated surface is set to the cathode potential, the ions will be accelerated due to the sheath potential drop before colliding with the surface. This acceleration phase would deviate the EDF of ions comparing to the bulk plasma. Motivation of this work is to measure the EDF of colliding ions with the surface of cathode. The GEA[5] is appropriate for discriminating ions with different energies while they are approaching the cathode surface.

Different types of gridded energy analyzers can be designed in order to detect and determine the energy distribution of ions or electrons. Energy analyzers may have one, double, or triple grids [21], [22], [23], [24]. In this work a single gridded energy analyzer was designed and constructed and its collector is connected to the cathode in order to determine the energy distribution of colliding ions with the cathode in dc glow discharge plasmas of argon.

This article is organized as following. Section II demonstrates the theory of the work. In section III simulation of the GDP is discussed. Section IV presents the simulation of the GEA in action and the steps of its design and construction. The experimental setup is explained in section V. Finally, the results and conclusions are provided in section VI and VII.

---

[4] Energy distribution function
[5] Gridded Energy Analyzer

## II. Theoretical basis and methods

In order to measure the EDF of colliding ions with the cathode surface (as a PFM), a single gridded energy analyzer has been considered while its collector is set to the cathode potential. Therefore, the approaching ions onto the collector would have similar EDF to the ions colliding with the cathode surface. Those ions that collide with the grid body will be eliminated from the measurement process. Depending on the grid transparency, which is related to the mesh size of the grid, a fraction of the approaching ions can pass through the grid and enter the space between the grid and the collector. If the ions reach the collector, their current is measured.

A schematic of the GEA in a plasma chamber is shown in Figure 1 (The chamber consists of two electrodes which are separated and held by four identical polyamide rods).

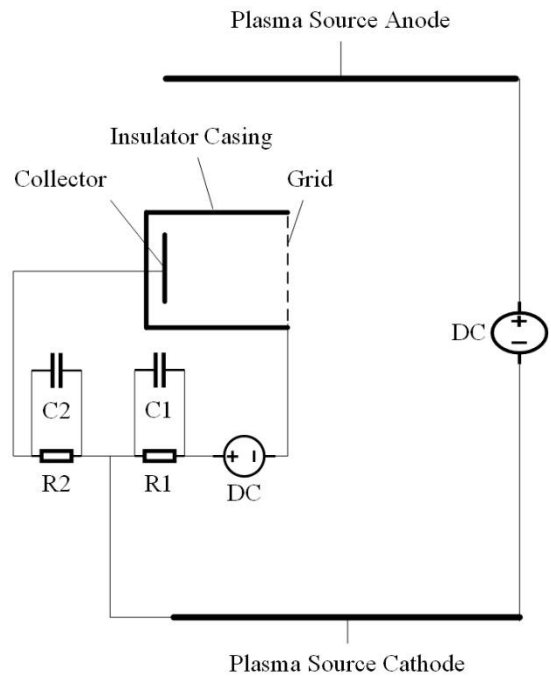

**Figure 1** Schematic of the GEA inside a dc glow discharge plasma chamber (The applied voltage between the chamber electrodes is 500 V or 600 V)

If the grid is biased to a certain level of negative potential with respect to the collector, then the ions with energies less than the potential barrier between the grid and the collector will be prevented from reaching the collector. The more the negative voltage is applied to the grid with respect to the collector, the fewer the number of ions can reach the collector. The current (the number of ions collided with the collector per second) at a certain negative voltage applied to the grid, provides the EDF of ions.

In theory, the EDF of plasmas in equilibrium is described by Maxwellian distribution, given by Eq. (1).

$$F(\varepsilon)=\frac{2}{(kT)^{3/2}}\sqrt{\frac{\varepsilon}{\pi}}\exp\left(\frac{-\varepsilon}{kT}\right) \quad (1)$$

Where $F(\varepsilon)$ is the Maxwellian EDF and k, T and $\varepsilon$ are the Boltzmann constant, the temperature and the energy, respectively [25].

In the cases of weak equilibrium, a common distribution function is the Druyvesteyn EDF in which the ion temperature is lower than the electron temperature and it is written in the form of Eq. (2) [26].

$$F_D(\varepsilon)=\frac{0.5648 n_e}{(kT)^{3/2}}\sqrt{\varepsilon}\exp\left(-0.243\left(\frac{\varepsilon}{kT}\right)^2\right) \quad (2)$$

$F_D(\varepsilon)$ points the Druyvesteyn EDF and $n_e$ is the electron density. In glow discharge plasmas at pressures around 0.066 Pa, the Maxwellian energy distribution is valid and for higher pressures in the order of 1.33 Pa the Druyvesteyn EDF dominants [12], [26].

Considering relatively low electron temperature in argon glow discharge plasmas, the ions are supposed to be just once ionized (i.e., $n_i = n_e$). Therefore, Eq. (2) can be used to estimate the ion energy distribution function [13].

By changing the potential (V) of the grid with respect to the collector (See Figure 1), the GEA provides an I-V curve where, I is the current between the collector and the grid which is derived due to collisions of the passing ions through the grid. The relation between the I-V curve and EDF of ions can be found as Eq. (3) and Eq. (4) considering Maxwellian or Druyvesteyn distributions, respectively [11], [27].

$$f_M(V) = \frac{1}{Aen} \frac{dI}{dV} \quad (3)$$

$$f_D(V) = \frac{m^2}{2\pi e^3 A} \frac{d^2I}{dV^2} \quad (4)$$

Here, A is the surface area of the electrode (here, it is the collector) facing ions or electrons.

## III. Plasma simulation

Argon glow discharge plasma was simulated by COMSOL Multiphysics to estimate plasma parameters. The plasma chamber is a cylinder with diameter of 23 cm and the high of 40 cm. Electric potentials of 500 V or 600 V were applied to the electrodes at the pressure of 5.33 Pa. Time-dependent study has been considered for the problem, and the simulation time is 1 second to obtain stable results. The simulation results have been presented in Figure 2 and Figure 3. The maximum values of the electron density between electrodes are $n_e = 1.46 \times 10^{16}$ m$^{-3}$ and $n_e = 2.24 \times 10^{16}$ m$^{-3}$ for 500 V and 600 V, respectively. While the corresponding peak values of the electron temperature profiles appear at the sheath close to the cathode surface ($T_e = 49$ eV and $T_e = 53$ eV, for 500 V and 600 V, respectively). The plasma potential profiles are quite uniform in the chamber except the sheath region on the cathode. Presence of a steep potential drop, which in principle can accelerate the bulk plasma ions toward the cathode surface are confirmed. This confirms the need to measure the EDF of the colliding ions with the cathode surface, which would be different from the bulk plasma.

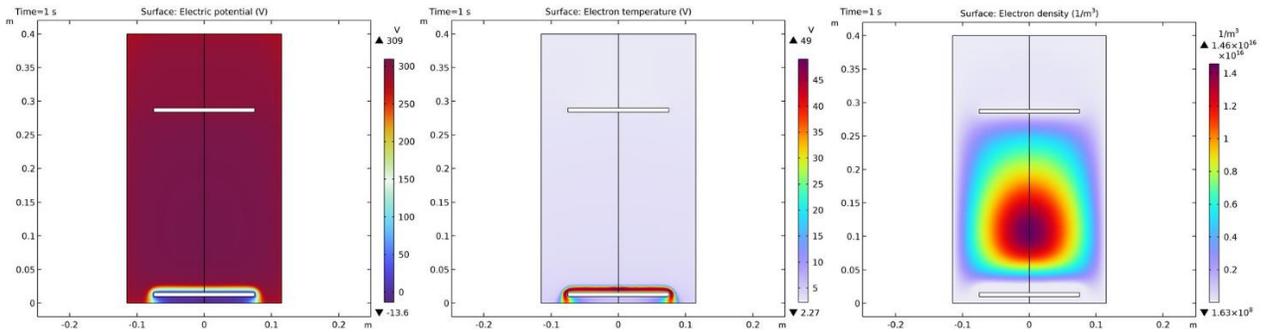

**Figure 2** Electric potential, electron temperature and electron density profiles of argon dc glow discharge plasma for discharge voltage of 500 V and gas pressure of 5.33 Pa

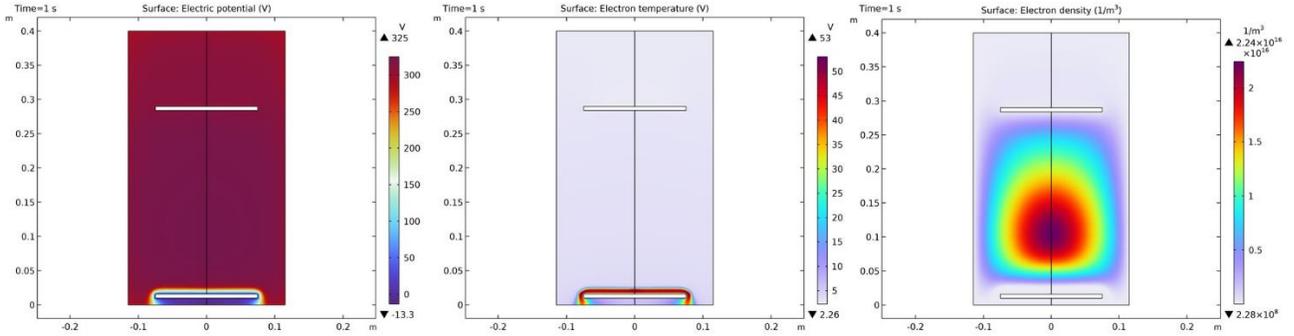

**Figure 3** Electric potential, electron temperature and electron density profiles of argon dc glow discharge plasma for discharge voltage of 600 V and gas pressure of 5.33 Pa

Figure 4 represents the number density versus the energy of electrons and so provides an approach of EDF for discharge voltages of 500 V and 600 V in argon plasmas at the pressure of 5.33 Pa. It is revealed from the figure that the maximum value of the number density is increased by increasing the discharge potential and it is also obvious from the tail of the EDF that the maximum energy of electrons is near 10 eV larger when the applied voltage reaches 600 V.

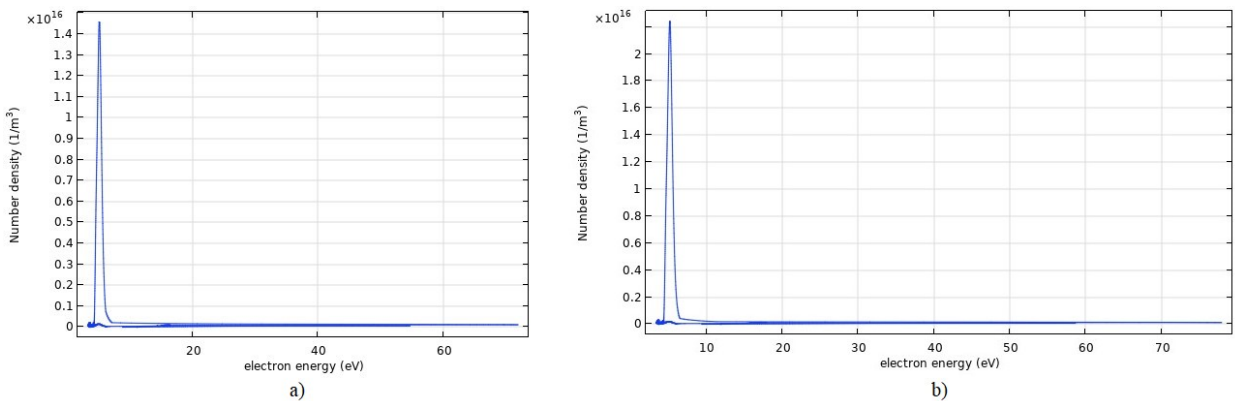

**Figure 4** Number density vs. the energy of electrons of argon dc glow discharge plasma at pressure of 5.33 Pa for the discharge potentials of a) 500 V and b) 600 V

Since the problem has been solved at thermodynamic equilibrium, the ion density and temperature are supposed to be close to the electron density and temperature. Consequently, the EDF of the bulk plasma ions is considered to be similar to the EDF of the electrons. Although these assumptions are not fully satisfied in low density dc glow discharge plasmas, but it can provide a platform in order to compare the EDF of the bulk plasma with that of ions that are accelerated due to the potential drop in the cathode sheath region [13], [28].

Therefore, a GEA has been considered in this work to experimentally measure the EDF of the colliding ions with the cathode surface. Admittedly, the information about the EDF of plasma is beneficial to reach an approach of the plasma behavior and the impacts on the PFMs.

## IV. Design and construction of the GEA

Since the GEA consists of a grid, the profile of the electric potential of the grid has been simulated as the first step. The grid size is considered to be 20 mm×20 mm×0.5 mm, and the area of each grid hole is 500 $\mu m^2$. The potential variation across the normal axis of each hole of the grid has been shown in Figure 5 (a). The figure indicates that at the distances of about 30 mm from the grid, the potential downgrades to zero.

Another essential issue is the electric potential variation on the surface of each grid hole. Figure 5 (b) represents the simulated potential profile across the surface of a hole with the diameter of 0.1 mm. The figure clearly shows that the electric potential is almost uniform over the hole surface at this size and the potential drop at the center of the grid holes is less than 1 %. It should be noted that in experiments if the hole diameter is larger than the Deby length of the plasma, the potential rapidly decreases by moving toward the center of the hole. Therefore, the hole size is limited by the Deby length of the plasma.

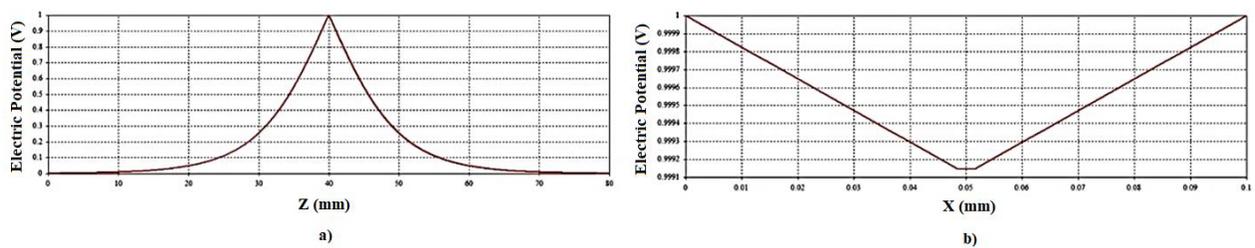

**Figure 5** Normalized profiles of electric potential: a) across the normal axis of a hole of the grid, and b) over the hole diameter

The grid and collector were supposed to be two discs with 1.25 cm of radiuses. The distance between the grid and collector is 20 mm. The grid thickness is 0.1 mm and it consists of 0.3×0.3

mm$^2$ square holes. Figure 6 shows the simulated potential variation around the grid and the collector with applied voltages of -30 V and 0, respectively.

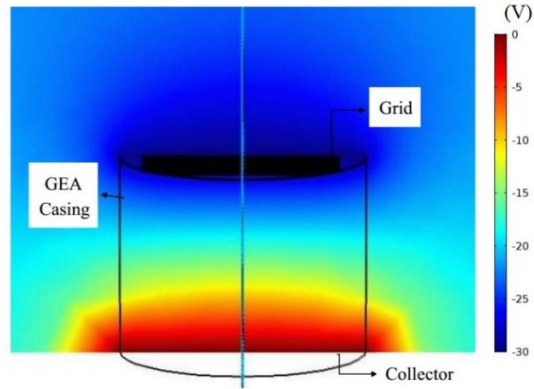

**Figure 6** Potential variation around the grid when it is set at -30 V and the collector is grounded

Finally, the simulation of the GEA in a chamber of mono energetic Ar$^+$ ions, with the energy of 5 eV, was considered. The grid is again at -30 V and the collector is grounded. Figure 7 presents the trajectories and energies of the originally mono energetic ions. It is clear that some ions have passed through the grid but they haven't had enough energy to reach the collector. Therefore, they've come back to the grid. The red bulbs are the ions which are more energetic than the others but they have collided with the grid surface and stopped.

From the results, 12 % of all ions can pass the grid and reach the collector. These ions have escaped from the attraction force due to the electric field of the grid and have lost some of their energy when they have reached the collector. This is why their energy is lower than those that hit the grid surface.

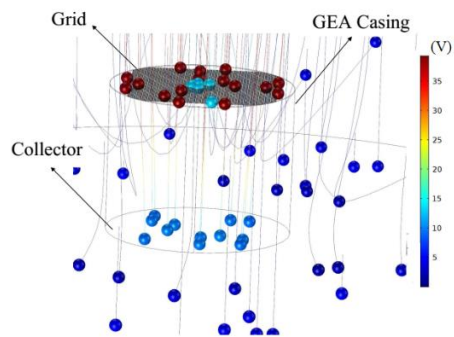

**Figure 7** Argon ions trajectories in the GEA which has been installed in a chamber of ions. The grid potential is -30 V, the collector is grounded and energy of all ions are supposed to be 5 eV

By considering the electron density and temperature and also the gas pressure, the Deby length is found to be in the order of 100 micrometers [13], [28], [29].

$$\lambda_{De} = \left(\frac{\varepsilon_0 T_e}{e^2 n_0}\right)^{1/2} \quad (5)$$

$$\lambda_{De} = 2.35 \times 10^{-5} \left(\frac{T_k}{n_{20}}\right)^{1/2} \text{ (m)} \quad (6)$$

$n_0$ is the unperturbed density of ions and electrons. $T_k$ is in keV, and the unit of $n_{20}$ is $10^{20}$ m$^{-3}$.

To sum up, the optimized width of the holes for the grid has been chosen 100 micrometers. The GEA is designed to have a cylindrical casing with 35 mm of radius and 5 cm of height. The device elements should be of sufficient strength against heat and tensions and should also be capable for machining. Therefore, the GEA casing was made of polyamide, the collector material is Aluminum and the grid is stainless steel.

The GEA needs an appropriate power supply by which the negative voltage can be swept from zero to an amount which provides a potential barrier by the grid to filter all ions. In order to determine magnitude of the power supply's voltage, the temperature of ions in the dc glow discharge plasma have been taken into consideration. Since the average electron energy in dc glow discharge plasma is between 2-5 eV, a power supply with voltage magnitude more than 10 times greater than the electron temperature can, in principle, filter all the ions. It should be noted that plasma ions and electrons have energy distribution which may include a few particles with energies much greater than the average energy. Eventually an adjustable 0 to -75 V dc power supply was used, while in the experiments the values below -50 V were not required [13], [28].

Considering the results of simulations, different distances between the grid and collector were examined and it was revealed that at higher distances the signal to noise ratio downgrades and for small distances, sparks and arcs between the electrodes result in noisy signals. Therefore, the collector was placed at 5 mm from the grid which is the optimized distance.

For more precise design, the mean free path of the plasma particles is calculated as following,

$$\text{mfp} = \frac{1}{\sqrt{2}n_B\sigma} \quad (7)$$

Where $n_B$ and $\sigma$ are the particle density and the cross section of collision between two particles, respectively (here, particle can be atom, molecule, ion, or electron) [30].

Considering the density range of dc glow discharge plasma and the collision cross section of Argon gas particles, from the Eq. (7) it is be revealed that the mean free path of gas in this work is so greater than the GEA dimensions that the interactions of ions inside the device can be overlooked (the mean free path is in the order of $10^5$ meters) [11], [29], [31].

A fact that should be considered for utilization of the GEA is the emission of secondary electrons. Plasma ions collide with the surface of the grid with probability of P and the secondary electrons are emitted from both sides of the grid. The electrons that are emitted from the back side of the grid (1/2 of all secondary electrons) can reach the collector and affect the output signal.

P = [(Total area of the grid) – (area of the grid holes)] / (Total area of grid)    (8)

For the grid that applied in this work, the collision probability of P was calculated to be 60 %. If the ions that collide to the grid surface are energetic enough to dominant its work function (W), they can result in the secondary electrons production. In this work, W is 4.4 eV for the grid and 4.28 eV for the collector. If the energy distribution function is f(E), the parameter g is defined as,

$$g = W \times f(E) \quad (9)$$

And the number of secondary electrons produced due to the collision of ions with the grid ($N_g$) is calculated as following.

$$N_g = (1/2) \times P \times g \quad (10)$$

Secondary electrons can also be emitted from the collector surface. The electrons emitted from both sides of the collector can affect the output signal and the surface has no hole. Therefore, the number of secondary electrons emitted from the collector is $N_c = g$.

Finally,

$$N_g = 1.32 f(E) \tag{11}$$

$$N_c = 4.28 f(E) \tag{12}$$

Since the mean free path of the secondary electrons is much larger than the distance between the grid and the collector and assuming local thermal equilibrium between the ions and the secondary electrons, the EDF of the secondary electrons is approximated to be equal to the EDF of ions. Therefore, the effect of the secondary electrons on the EDF of the ions can be estimated as following (The secondary electrons which emit from the grid increase the output signal and the secondary electrons emit from the collector diminish the signal).

$$f_{Ex}(E) = f_{Re}(E) - 1.32 f_{Re}(E) + 4.28 f_{Re}(E) \tag{13}$$

$$f_{Re}(E) = 0.252 f_{Ex}(E) \tag{14}$$

Where $f_{Ex}(E)$ represents the measured EDF from experiment and $f_{Re}(E)$ is the real energy distribution function.

To sum up, the effect of the production of the secondary electrons was considered by the calculated coefficient factor, which indicates that the actual distribution function is just linearly proportional to the experimental distribution function. Therefore, the experimental distribution can be corrected in order to compensate the secondary electrons contribution and the shape of the actual distribution should be roughly similar to the experimental distribution. Hence, although the additional grid hasn't been used in this work for secondary electrons, the effect of them has been taken into consideration using some assumptions.

## V. Experimental setup

All experiments have been performed in a cylindrical vacuum chamber with diameter of 23 cm and height of 40 cm. Two identical movable Aluminum discs with diameters of 15 cm were used as the discharge electrodes. A DC power supply (0 to 2 kV, 2 A) was used and two values of 500 V and 600 V potentials were applied to produce plasmas. The Argon gas pressure has been controlled by a Pirani Vacuum Gauge and a needle valve. The gas pressure was kept at 5.33 Pa for all tests. To measure I-V curves, another adjustable dc power supply has been used to negatively bias the grid with respect to the collector while the collector itself is connected to the cathode electrode of the plasma chamber.

## VI. Results

In this work, the EDF of ions has been investigated to reach an approach about plasmas effects on the PFMs considering different applications of the GDP. To achieve this objective, a GEA has been designed and constructed to experimentally measure the EDF of argon ions colliding with the cathode surface. Figure 8 shows the experimental I-V curve when the plasma chamber electrodes are set to a potential difference of 500 V at the pressure of 5.33 Pa for argon gas.

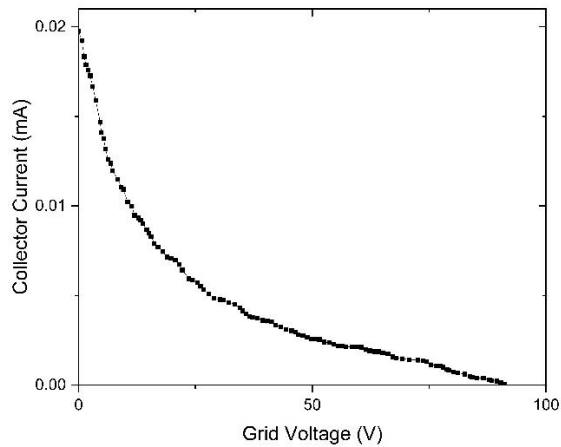

**Figure 8** The current of ions approach to the collector versus the potential applied to the grid at 5.33 Pa for discharge potential of 500 V

Figure 9 presents the experimental EDF of colliding ions with the cathode for discharge voltages of 500 V and 600 V. The obtained EDFs contain three or four distinguished peaks.

The presence of multiple peaks in the EDF of ions means that their population consists of different groups of ions and each group is in equilibrium with its own temperature. The simulated EDF of ions which have been discussed in the Sec. III can be compared with the experimental EDF. It is clear that in the simulated EDFs there is one peak that can be modeled by the Druyveteyn EDF. In order to perform comparison, a Druyveteyn EDF is fitted to the first peak of the experimental EDF and the corresponding value of the ion temperature is calculated. The obtained values of the ion temperature for discharge voltages of 500 V and 600 V are respectively 3.5 eV and 4 eV.

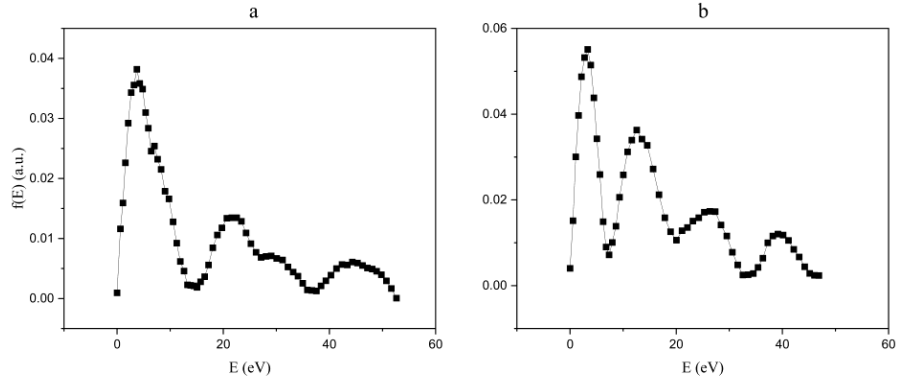

**Figure 9** Normalized EDFs of colliding argon ions with the cathode electrode in dc glow discharge plasma at the pressure of 5.33 Pa for discharge potentials of a) 500 V, and b) 600V

According to Figure 9 and Figure 4, the most probable energy of ions, which correspond to the first hump, is in reasonable agreement with those of the simulations results. But in the experimental EFDs, considerable number of ions are distributed from 10 eV up to 50 eV. The peak of the first humps in the experimental EDFs and also the peak of the simulated EDFs, both rise by increasing the discharge voltage but on the other hand in the case of the higher discharge voltage the difference between the principal hump and the second hump decreases with respect to the lower discharge voltage. This shows by increasing the discharge voltage, the EDF tends toward higher energies which is in agreement with the theoretical and simulated EDFs but the way the EDF moves toward high energies is different from what simulation and theory predict. Of course, one of the main reasons could be the fact that the thermodynamic equilibrium is not completely satisfied in experiment. The presented work can provide a new insight to understand the way the plasma ions have arranged themselves while they are accelerated in the cathode sheath region. Also, measurement of the experimental EDF of colliding ions with the surface of the cathode would help to improve methods to control the flux and energy of ions in a vast varieties of plasma sources and investigate the effects on the PFMs.

## VII. Conclusion

GDP is a widely utilized technique for applications, including dust removal from the PFMs in Tokamaks. The EDF of plasma ions provides useful information about plasma parameters and helps to reach an insight about the plasma effects on the PFMs. To measure the EDF, a GEA has been designed and constructed and then installed in a plasma chamber (the collector of the GEA is connected to the cathode of the plasma chamber). The EDF of plasma ions impinging the surface of the cathode has been experimentally measured and the following results were obtained:

- ✓ The EDFs comprise multiple distinguished peaks. This means that the experimental plasma had included different groups of ions, the particles of each group are in equilibrium, and have different temperature compared to that of other groups.
- ✓ The first humps in the experimental EDFs are comparable to the corresponding EDFs from the simulation.
- ✓ By increasing the discharge potential of plasma, the peak of first hump in the experimental EDF rises, which is in agreement with the rise of the simulated EDF peak.
- ✓ By increasing the discharge potential of plasma, the distance between the principal hump and the second hump decreases. It is revealed that at higher energies, the plasma groups interact with each other and thermodynamic equilibrium is going to be more satisfied comparing to lower discharge potentials. While in the simulation, thermodynamic equilibrium has been considered to be completely satisfied.